\documentstyle[prb,aps,twocolumn,epsfig]{revtex}
\begin{document}

\preprint{\hfill  MMM99  CD-01}

\title{({\it Invited}) Landau Zener method to study quantum phase interference\\ 
of Fe$_8$ molecular nanomagnets
}

\author{W.~Wernsdorfer$^1$, R.~Sessoli$^2$, A.~Caneschi$^2$, D.~Gatteschi$^2$, A.~Cornia$^3$, D.~Mailly$^4$}
\address{
$^1$Lab. Louis N\'eel, associ\'e \`a l'Universit\'e Joseph Fourier, 
CNRS, BP 166, 38042 Grenoble Cedex 9, France\\
$^2$Department of Chemistry, University of Firenze, via Maragliano 72, 50144 
Firenze, Italy\\
$^3$Department of Chemistry, University of Modena, via G. Campi 183, 41100 Modena, Italy\\
$^4$Lab. de Microstructures et de Micro\'electronique, 196 av. H. Ravera, 92220 
Bagneux, France
}

%

%

\date{\today} \maketitle

\begin{abstract}
We present details about an experimental method based on the Landau Zener model which 
allows to measure very small tunnel splittings $\Delta$ in molecular clusters Fe$_8$. The 
measurements are performed with an array of micro-SQUIDs. The observed oscillations 
of $\Delta$ as a function of the magnetic field applied along the hard anisotropy axis are 
explained in terms of topological quantum interference of two tunnel paths of opposite 
windings. Transitions between $M = -S$ and $(S - n)$, with $n$ even or odd, revealed a 
parity (symmetry) effect which is analogous to the suppression of tunneling predicted for 
half integer spins. This observation is the first direct evidence of the topological part of the 
quantum spin phase (Berry phase) in a magnetic system. The influence of intermolecular 
dipole interactions on the measured tunnel splittings $\Delta$ are shown.
\end{abstract}
\bigskip
\narrowtext

\section {Introduction}
Magnetic molecular clusters are the final point in the 
series of smaller and smaller magnets 
from bulk matter to atoms. Up to now, they have been the most promising candidates to 
observe quantum phenomena since they have a well defined structure with well 
characterized spin ground state and magnetic anisotropy. These molecules are regularly 
assembled in large crystals where often all molecules have the same orientation. Hence, 
macroscopic measurements can give direct access to single molecule properties. The most 
prominent examples are a dodecanuclear mixed-valence manganese-oxo cluster with 
acetate ligands, Mn$_{12}$ acetate~\cite{Sessoli93}, and an octanuclear iron(III) oxo-
hydroxo cluster of formula [Fe$_8$O$_2$(OH)$_{12}$(tacn)$_6$]$^{8+}$, 
Fe$_8$~\cite{Barra96}, where tacn is a macrocyclic ligand. Both systems have a spin 
ground state of $S = 10$, and an Ising-type magneto-crystalline anisotropy, which 
stabilizes the spin states with $M = \pm10$ and generates an energy barrier for the 
reversal of the magnetization of about 67~K for Mn$_{12}$ acetate and 25~K for 
Fe$_8$.

Fe$_8$ is particular interesting because its magnetic relaxation time becomes temperature 
independent below 0.36~K showing for the first time that a pure tunneling mechanism 
between the only populated $M = \pm10$ states is responsible for the relaxation of the 
magnetization~\cite{Sangregorio97}. Measurements of the tunnel splitting $\Delta$ as a 
function of a field applied in direction of the hard anisotropy axis showed oscillations of 
$\Delta$, i.e. oscillations of the tunnel rate~\cite{Science}. In a semi-classical 
description~\cite{Berry84,Haldane88}, these oscillations are due to constructive or 
destructive interference of quantum spin phases of two tunnel paths~\cite{Garg93,others}. 
Furthermore, parity effects were observed when comparing the transitions between 
different energy levels of the system~\cite{Science} which are analogous to the 
parity effect between systems with half integer or integer spins~\cite{Loss92,Delft92}. An 
alternative explication in terms of intermediate spin was also presented~\cite{Barnes}. 
Hence, molecular chemistry had a large impact in the research of quantum tunneling of 
magnetization at molecular scales.

\begin{figure}[t]
\centerline{\epsfxsize=7.5 cm \epsfbox{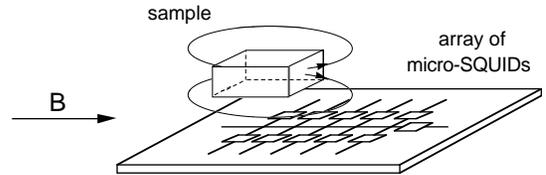}}
\caption{Schematic representation of our magnetometer which is an array of micro-
SQUIDs. Its high sensitivity allows us to study single crystals of the order of 10 to 
500~$\mu$m which are placed directly on the array.}
\label{SQUID_array}
\end{figure}

\section{Micro-SQUID technique}
The technique of micro-SQUIDs is very similar to the traditional SQUID technique. The 
main difference is that the pick-up coil is replaced by a direct coupling of the sample with 
the SQUID loop (Fig.~\ref{SQUID_array}). When a small sample is directly placed on 
the wire of the SQUID loop, the sensitivity of the micro-SQUID technique is ten orders 
of magnitude better than a traditional SQUID \cite{WW_PhD} reaching $10^{-17}$ 
emu. This sensitivity is smaller when the sample is much bigger than the micro-SQUID. 

Our new magnetometer is a chip with an array of micro-SQUIDs. The sample is placed 
on top of the chip so that some SQUIDs are directly under the sample, some SQUIDs are 
at the border of the sample and some SQUIDs are beside the sample 
(Fig.~\ref{SQUID_array}). When a SQUID is very close to the sample, it is sensing 
locally the magnetization reversal whereas when the SQUID is far, it is integrating over a 
bigger sample volume.

The high sensitivity of this magnetometer allows us to study single Fe$_8$ crystals 
\cite{Wieghardt} of the order of 10 to 500 $\mu$m. The magnetometer works in the 
temperature range between 0.035 and 6~K and in fields up to 1.4~T with sweeping rates 
as high as 1~T/s, and a field stability better than a microtesla. The time resolution is about 
1~ms allowing short-time measurements. The field can be applied in any direction of the 
micro-SQUID plane with a precision much better than 0.1$^{\circ}$ by separately 
driving three orthogonal coils \cite{WW_PhD}. In order to ensure a good thermalisation, 
the crystal is fixed by using a mixture of araldite and silver powder. 

\begin{figure}
\centerline{\epsfxsize=7.5 cm \epsfbox{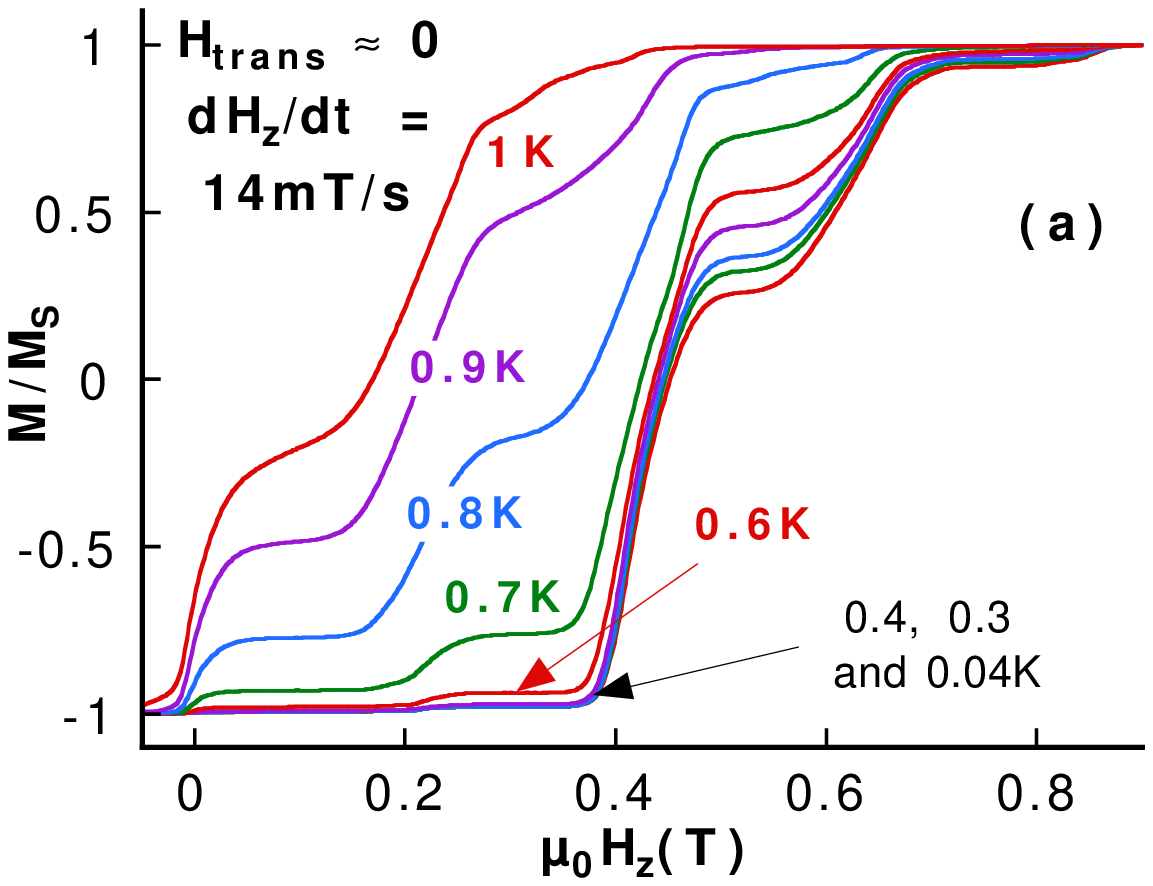}}
\centerline{\epsfxsize=7.5 cm \epsfbox{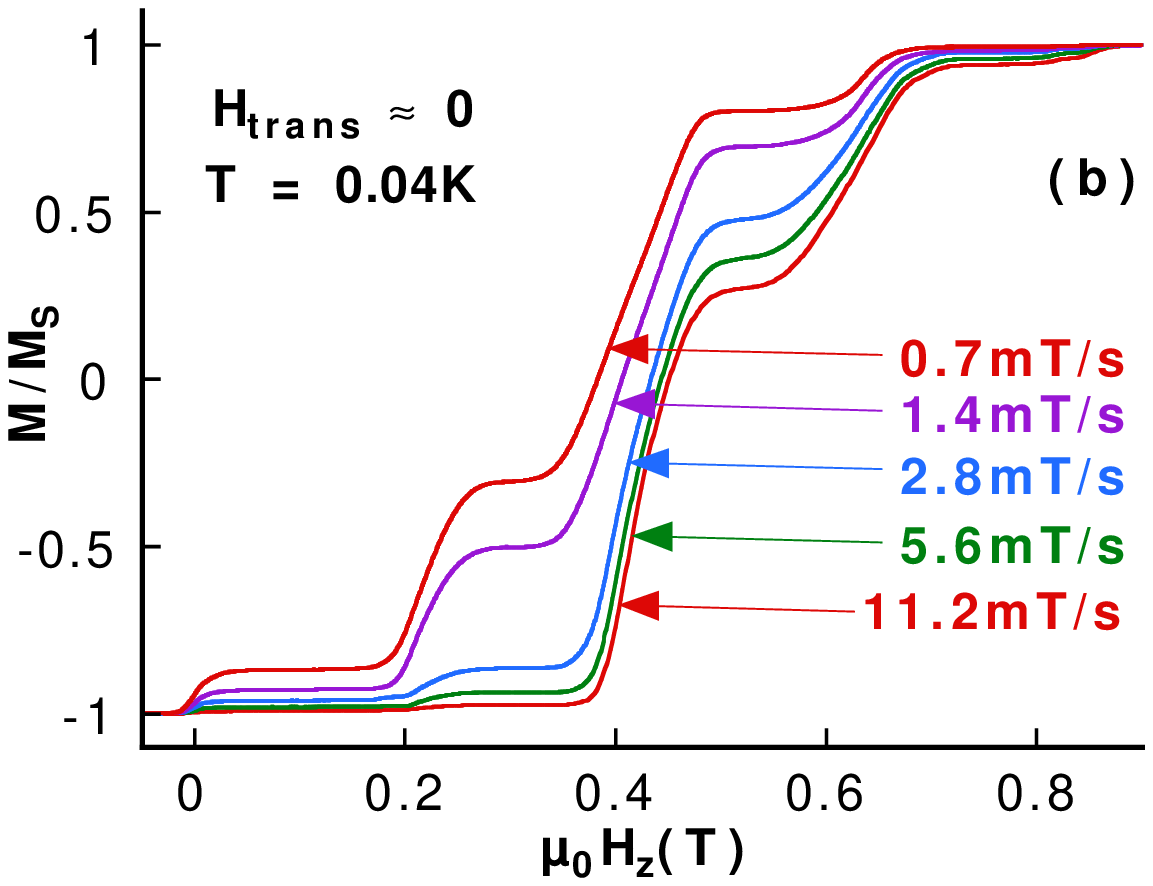}}
\caption{Temperature (a) and field sweeping rate (b) dependence of hysteresis loops of Fe$_8$ molecular clusters \cite{remark_heat}.
Resonant tunneling is evidenced by equally separated steps of $\Delta H_z \approx$ 
0.22~T which, at $T <$ 360~mK, correspond to tunnel transitions from the state $M = 
-10$ to $M = 10 - n$, with $n = 0, 1, 2...$. The resonance widths of about 0.05~T are 
due to mainly dipolar fields between the molecular 
clusters~\cite{Ohm98,PRL_dig}.}
\label{fig_hyst}
\end{figure}

\begin{figure}
\centerline{\epsfxsize=7.5 cm \epsfbox{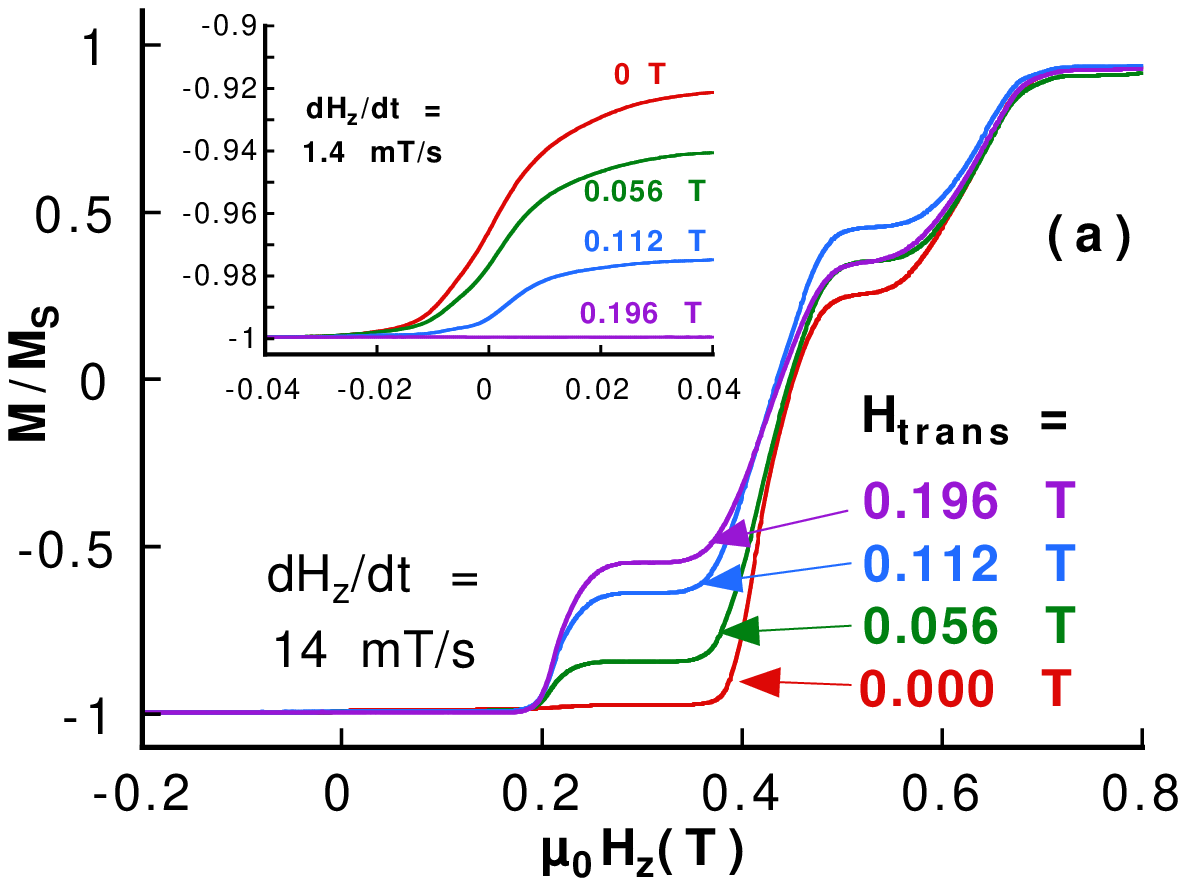}}
\centerline{\epsfxsize=7.5 cm \epsfbox{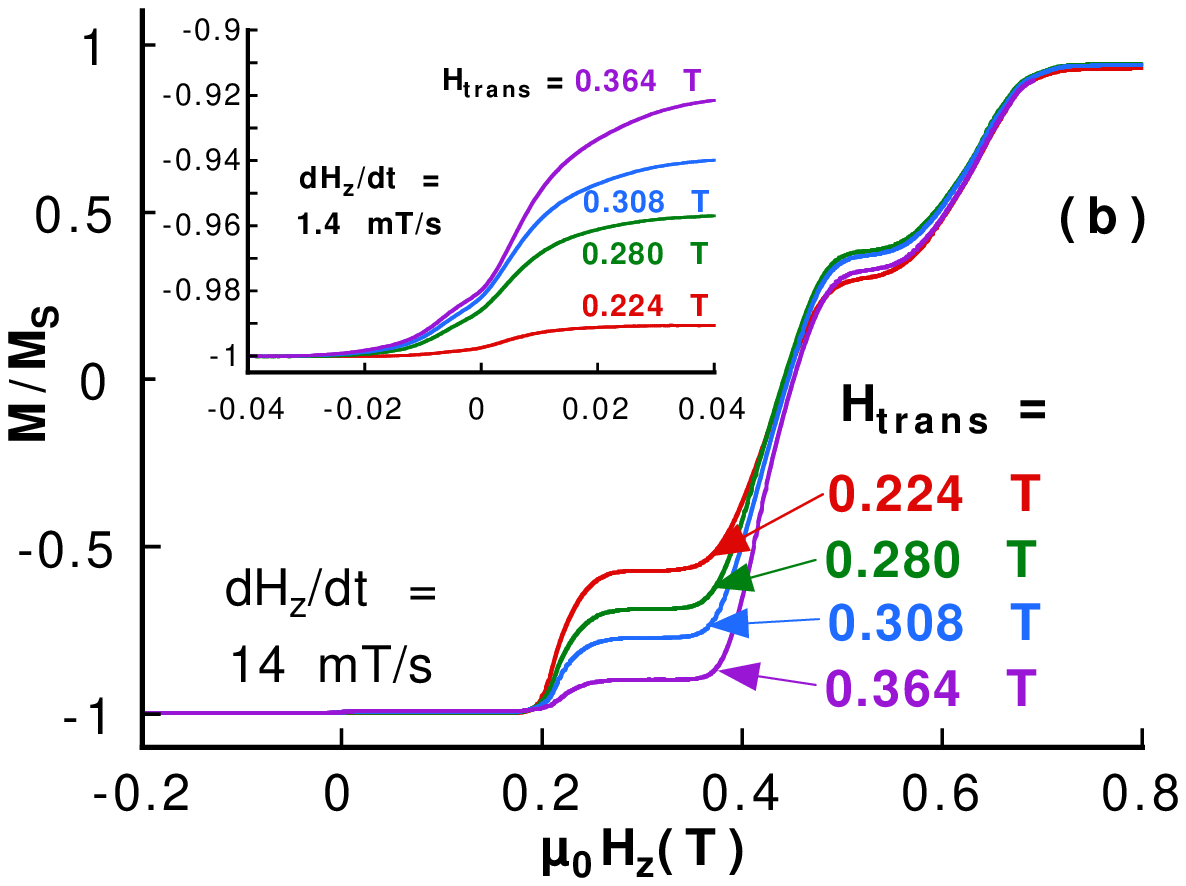}}
\caption{Hysteresis loops measured in the presence of a constant 
transverse field, at 0.04 K. 
Insets: enlargement around the field $H = 0$. 
Notice that the sweeping rate is ten times slower 
for the measurements in the insets than that of the main figures.}
\label{fig_hyst_Hx}
\end{figure}

Typical measurements of  magnetic hysteresis curves for a crystal of molecular Fe$_8$ 
clusters are displayed in Fig. \ref{fig_hyst} and \ref{fig_hyst_Hx}. 
The field was swept in direction of the easy axis of 
magnetization evidencing about equally separated steps at 
$H_z \approx n \times 0.22$~T ($n =$ 1, 2, 3 ...)
which are due to a faster relaxation of magnetization at particular field values. The step 
heights (i.e. the relaxation rates) change when a constant transverse field is applied. It is 
the purpose of these article to present a detailed study of this behavior which is 
interpreted in terms of resonant tunneling between discrete energy levels of the spin 
Hamiltonian $S = 10$ of Fe$_8$.

\section{Landau Zener method}
\begin{figure}[t]
\centerline{\epsfxsize=5.5 cm \epsfbox{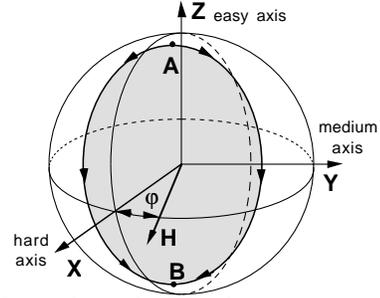}}
\caption{Unit sphere showing degenerate minima A and B which are joined by two tunnel 
paths (heavy lines). The hard, medium, and easy axes are taken in $x$, $y$ and 
$z$ direction, respectively. 
The transverse field $H_{trans}$ is applied in the $xy$ plane at an azimuth angle 
$\varphi$. At zero applied field, the giant spin reversal results from the interference of two 
quantum spin paths of opposite direction in the easy anisotropy plane $yz$.
By using Stokes theorem it has been shown \cite{Garg93} that 
the path integrals can be converted in an area integral, 
giving that destructive interference, that is a quench of the tunneling rate, occurs whenever 
the shaded area is $k \pi / S$, where $k$ is an odd integer.}
\label{sphere}
\end{figure}

\begin{figure}[t]
\centerline{\epsfxsize=7.5 cm \epsfbox{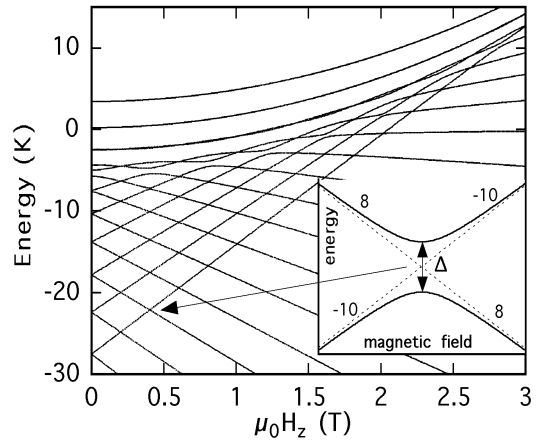}}
\caption{Zeeman diagram of the 21 levels of the $S = 10$ manifold of Fe$_8$ as a 
function of the field applied along the easy axis (\ref{eq_H_biax}). 
Form the bottom to the 
top, the levels are labeled with quantum numbers $M = \pm10, \pm9, ...0$. 
The levels cross 
at fields given by $\mu_0 H_n \sim n\ 0.22$T, with $n = 1, 2, 3 ...$. The inset displays 
the detail at a level crossing where the transverse terms (terms containing $S_x$ 
or/and $S_y$ spin operators) turn the crossing into an avoided crossing. The higher the gap $\Delta$, the stronger is the tunnel rate.}
\label{fig_level}
\end{figure}

The simplest model describing the spin system of Fe$_8$ molecular clusters has the 
following Hamiltonian \cite{Barra96}:
\begin{equation}
H = -D S_z^2 + E \left(S_x^2 - S_y^2\right) + g \mu_B \vec{S}\vec{H}
\label{eq_H_biax}
\end{equation}
$S_x$, $S_y$, and $S_z$ are the three components of the spin operator, $D$  and $E$ 
are the anisotropy constants, and the last term of the Hamiltonian describes the Zeeman 
energy associated with an applied field $H$. This Hamiltonian defines a hard, medium, 
and easy axes of magnetization in $x$, $y$ and $z$ direction, 
respectively (Fig.~\ref{sphere}). It 
has an energy level spectrum with $(2S+1) = 21$ values which, in first approximation, 
can be labeled by the quantum numbers $M~= -10, -9, ...10$. 
The energy spectrum, 
shown in Fig.~\ref{fig_level}, can be obtained by using 
standard diagonalisation techniques of the $[21 \times 21]$ 
matrix describing the spin 
Hamiltonian $S = 10$.
In the low temperature limit 
($T <$ 0.36~K) only the two lowest 
energy levels with $M = \pm10$ are occupied. The 
avoided level crossing around $H_z$ = 0 is due to transverse terms containing $S_x$ or 
$S_y$ spin operators (see inset of Fig.~\ref{LZ_test}). The spin $S$ is 'in resonance' 
between two states when the local longitudinal field is close 
to the avoided level crossing ($< 10^{-8}$ T for the avoided level crossing 
around $H_z$ = 0). The energy gap, the so-called 
tunnel spitting $\Delta$, can be tuned by an applied field in the $xy$--plane 
(Fig.~\ref{sphere}) via the $S_xH_x$ and $S_yH_y$ Zeeman terms. It turns out that a 
field in $H_x$ direction (hard anisotropy direction) can periodically change the tunnel 
spitting $\Delta$. In a semi-classical description, these oscillations are due to constructive 
or destructive interference of quantum spin phases of two tunnel paths 
(Fig.~\ref{sphere}). The period of oscillation is given by \cite{Garg93}:
\begin{equation}
\Delta H = \frac {2 k_B}{g \mu_B} \sqrt{2 E (E + D)}
\label{eq_Garg}
\end{equation}
The most direct way of measuring the tunnel splitting $\Delta$ is by using the Landau-
Zener model \cite{Zener,Zener_new} which gives the tunneling 
probability $P$ when sweeping the 
longitudinal field $H_z$ at a constant rate over the avoided energy 
level crossing (see inset of Fig.~\ref{fig_level}):
\begin{equation}
P_{M,M'} = 1 - e^{-
\frac {\pi \Delta_{M,M'}^2}{2 \hbar g \mu_B |M - M'| dH/dt}}
\label{eq_LZ}
\end{equation}
Here, $M$ and $M'$ are the quantum numbers of the avoided 
level crossing, $dH/dt$ is the 
constant field sweeping rates, $g~\approx~2$, $\mu_B$ the Bohr magneton, and $\hbar$ 
is Planck's constant. In the following, we drop the index $M$ and $M'$. For very small tunneling probabilities $P$, we did multiple sweeps 
of the resonance transition. The relaxed magnetization after 
$N$ sweeps is given by ($n = 0$):
\begin{equation}
M(N) \sim exp\left\lbrack -2 P N \right\rbrack = exp\left\lbrack -\Gamma t \right\rbrack
\label{eq_M_N}
\end{equation}
where $N~= \frac{1}{A}\frac{dH}{dt} t$ is the number of sweeps over the level 
crossing, $\Gamma~= 2 P \frac{1}{A}\frac{dH}{dt} = \frac{\Delta M}{M_s} 
\frac{1}{A} \frac{dH}{dt}$ is the overall Landau-Zener 
transition rate, $\Delta M$ is the 
change of magnetization after one sweep, and $A$ is the amplitude of the ac-
field~\cite{remark1}. We have therefore a simple tool to obtain the tunnel splitting by 
measuring $P$, or $M(N)$ for $P << 1$.

\begin{figure}[t]
\centerline{\epsfxsize=8 cm \epsfbox{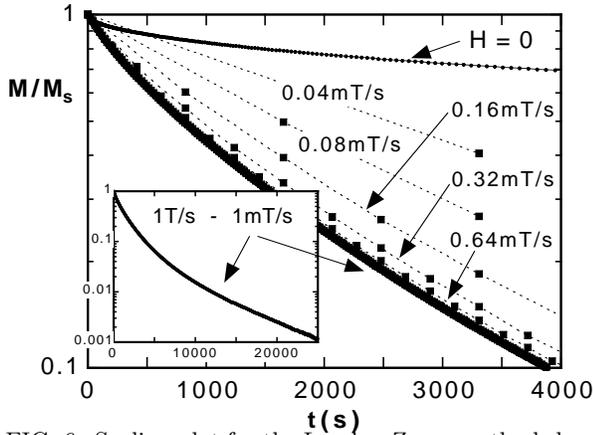}}
\caption{Scaling plot for the Laudau-Zener method showing the predicted field sweeping 
rate dependence for 1~T/s to 1~mT/s. Each point indicates the magnetization after a field sweep over the $M = \pm10$ resonance. The dotted lines are guides for the eyes. For comparison, there is also a relaxation curve 
without ac-field at $H = 0$. Inset: detail of the relaxation in the long time regime.}
\label{LZ_test_1}
\end{figure}

\begin{figure}[t]
\centerline{\epsfxsize=8 cm \epsfbox{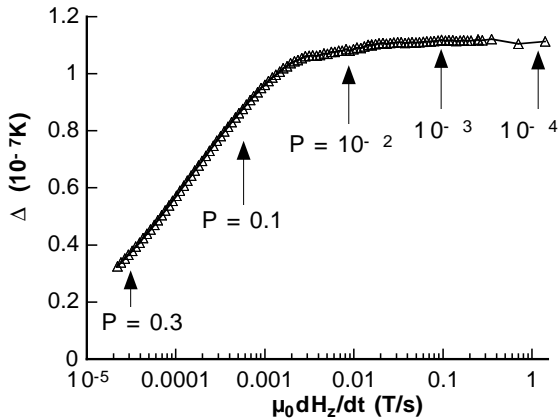}}
\caption{Filed sweeping rate dependence of the effective tunnel splitting $\Delta$ 
measured by a Landau Zener method which 
works in the region of high sweeping rates where $\Delta$ is sweeping rate independent. 
The measured Landau-Zener tunneling probability $P$ is indicated for 
several field sweeping rates.}
\label{LZ_test}
\end{figure}

In order to apply the Landau-Zener formula (Eq.~\ref{eq_LZ}), we first saturated the 
sample in a field of $H_z$~=~-1.4~T, yielding $M_{\rm in} = -M_s$ \cite{remark4}. 
Then, we swept the applied field at a constant rate 
over one of the resonance transitions and measured the fraction of molecules which 
reversed their spin. This procedure yields the tunneling rate $P$ and thus the tunnel 
splitting $\Delta$ (Eq.~\ref{eq_LZ}). We first checked the predicted Landau-Zener 
sweeping field dependence of the tunneling rate. This can be done by plotting the 
relaxation of magnetization as a function of $t = N \frac{A}{dH/dt}$. The Landau-
Zener model predicts that all measurements should fall on one line which was indeed the 
case for sweeping rates between 1 and 0.001~T/s (fig.~\ref{LZ_test_1}). The deviations 
at lower sweeping rates, are mainly due to the 'hole-digging mechanism'~\cite{PRL_dig} 
which slows down the relaxation \cite{remark5}. 
In the ideal case, we should find an exponential curve 
(Eq.~\ref{eq_M_N}). However, this might be only the case in the long time regime (see 
inset of Fig.~\ref{LZ_test_1})~\cite{remark2}. The origin is not clear but it might be 
related to dipolar interactions and hyperfine couplings. For comparison, there is also a 
relaxation curve without ac-field at $H = 0$ which shows much slow 
relaxation~\cite{PRL_dig}.

We also compared the tunneling rates found by the Landau-Zener method with those 
found using a square-root decay method which was proposed by Prokof'ev and 
Stamp~\cite{Prok_Stamp}, and found again a good agreement~\cite{PRL_dig}.

These measurements show that the Landau-Zener method is particularly adapted for 
molecular clusters because it works even in the presence of dipolar and hyperfine fields 
which spread the resonance transition provided that the field sweeping rate is not too 
small~\cite{remark3}.

\section{Oscillations of tunnel splitting}

\begin{figure}[t]
\centerline{\epsfxsize=8 cm \epsfbox{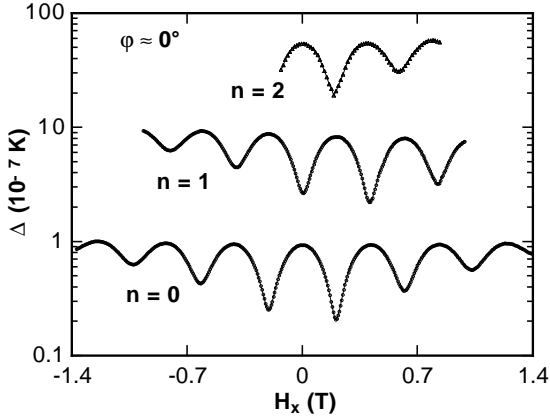}}
\caption{Measured tunnel splitting $\Delta$ as a function of transverse field for $\varphi 
\approx 0^{\circ}$, and for quantum transition between 
$M = -10$ and $(S - n)$. Note the parity 
effect which is analogous to the suppression of tunneling predicted for half integer spins 
\cite{Loss92}. It should also be mentioned that internal dipolar and hyperfine fields hinder a 
quench of $\Delta$ which is predicted for an isolated spin.}
\label{Delta_Hx}
\end{figure}

Studies of the tunnel splitting $\Delta$, at the tunnel transition 
between $M = \pm10$, as a function of transverse fields 
applied at different angles $\varphi$, defined as the 
azimuth angle between the anisotropy hard axis and the 
transverse field (Fig.~\ref{sphere}) show that 
for small $\varphi$ angles the tunneling rate oscillates with a period between minima of 
ca. 0.41~T, whereas no oscillations showed up for large $\varphi$ angles (see Fig.~2A in 
Ref.~\cite{Science}). In the latter case, a much stronger increase of $\Delta$ with 
transverse field is observed. The transverse field dependence of the tunneling rate for 
different resonance conditions between the state $M = -10$ and $(S - n)$ can be 
observed by sweeping the longitudinal field around $H_z~= n \times 0.22$~T with n~= 
0, 1, 2,~... \cite{remark4} The corresponding tunnel splittings $\Delta$ oscillate with almost the same 
period of ca. 0.4~T (Fig.~\ref{Delta_Hx}). In addition, comparing quantum transitions between $M = 
-S$ and $(S - n)$, with $n$ even or odd, revealed a parity (symmetry) effect which is 
analogous to the (Kramers) suppression of tunneling predicted for half integer 
spins~\cite{Loss92,Delft92}. This behavior has been observed 
for $n~=~0$ to~4 \cite{remark4}. A similar strong 
dependence on the azimuth angle $\varphi$ was observed for all the resonances.

\begin{figure}[t]
\centerline{\epsfxsize=8 cm \epsfbox{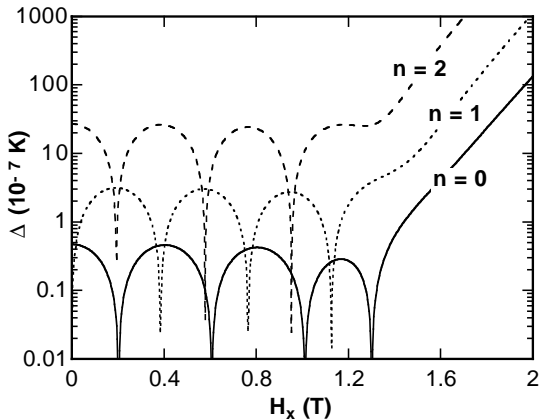}}
\caption{Calculated tunnel splitting $\Delta$ (Eq. 3) as a function of transverse field for 
quantum transition between $M = -10$ and $(10 - n)$ 
at $\varphi = 0^{\circ}$. Our choice of the forth order terms 
suppress the oscillations of $\Delta$ for $|Hx| > 1.4$ T.
Future measurements should focus on higher transverse fields.}
\label{Delta_Hx_theo}
\end{figure}

In the frame of the simple giant spin model (Eq. 1), the period of oscillation (Eq. 
2) is $\Delta H$~= 0.26~T for $D$~= 0.275~K  
and $E$~= 0.046~K as in Ref.~\cite{Barra96}. 
This is significantly smaller than the experimental value of ca. 0.4 T. In order to 
quantitatively reproduce the observed periodicity we have included forth order terms in the 
spin Hamiltonian (Eq.1) as recently employed in the simulation of inelastic neutron 
scattering measurements \cite{Caciuffo} and performed a 
diagonalization of the $[21 \times 21]$ 
matrix describing the S = 10 system. However, as the forth order terms are very small, 
only the term in $C(S_+^4 + S_-^4)$, which is the most efficient in affecting the tunnel 
splitting $\Delta$, has been considered for the sake of simplicity. The calculated tunnel 
matrix elements for the states involved in the tunneling process 
at the resonances $n$ = 0, 1, and 2 are reported in Fig. \ref{Delta_Hx_theo}, 
showing the oscillations as well as the parity effect for odd 
resonances. The period is reproduced  by using $D$ = 0.292 K  $E$ = 0.046 K as in Ref. \cite{Caciuffo}
but with a different $C$ value of $-2.9 \times 10^{-5}$ K. 
The calculated tunneling splitting is however  
ca. 3 times smaller  than the observed one \cite{remark3}. 
These small discrepancies are not surprising. 
In fact with the $C$ parameter we take into account the effects 
of the neglected higher order 
terms in $S_x$ and $S_y$ of the spin hamiltonian 
which, even if very small, can give an important 
contribution to the period of oscillation and dramatically 
affect $\Delta$, as first pointed 
out by Prokof'ev and Stamp \cite{Prok_Stamp}. 
Our choice of the forth order terms 
suppress the oscillations of $\Delta$ for $|Hx| > 1.4$ T which could not be studied in the current set-up. Future measurements should focus on the higher field region in order to find a better effective Hamiltonian.

\section{Intermolecular dipole interaction}

\begin{figure}[t]
\centerline{\epsfxsize=8 cm \epsfbox{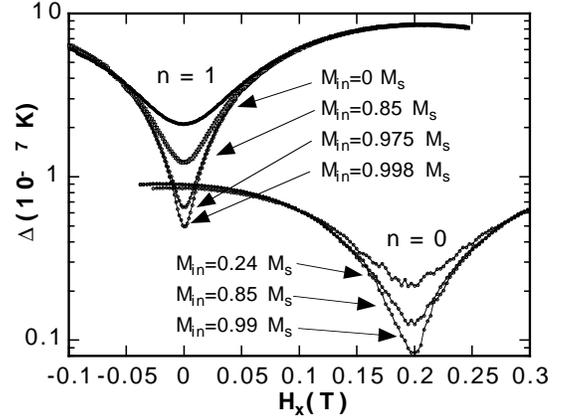}}
\caption{Detailed measurement of the tunnel splitting $\Delta$ around a 
topological quench for the quantum transition between 
$M = -10$ and $(10 - n)$ at $\varphi = 0^{\circ}$. 
Note the strong dependence on the initial magnetization 
which demonstrates the dipolar interaction between 
Fe$_8$ molecular clusters \cite{PRL_dig}.}
\label{Delta_Hx_Min}
\end{figure}

Fig.~\ref{Delta_Hx_Min} shows detailed measurement of the 
tunnel splitting $\Delta$ around a topological 
quench for the quantum transition between $M = \pm10$, and $M = -10$ and $9$. 
Particular effort were made to align well the transverse field in direction of the hard axis. 
The initial magnetizations $0 \leq M_{\rm in} \leq M_s$ were prepared by rapidly quenching 
the sample from 2~K in the present of an longitudinal applied field $H_z$. The quench 
takes approximately one second and thus the sample does not have time to relax, either by 
thermal activation or by quantum transitions, so that the high temperature ``thermal 
equilibrium'' spin distribution is effectively frozen in. For $H_z~>$~1~T, one gets an 
almost saturated magnetization state. 

The measurements of $\Delta(M_{in})$ show a strong dependence of the minimal tunnel 
spittings on the initial magnetization (Fig.~\ref{Delta_Hx_Min}) 
which demonstrates the transverse dipolar 
interaction between Fe$_8$ molecular clusters being largest of $M_{\rm in} = 0$ similar 
to the longitudinal dipolar interaction~\cite{PRL_dig}.

\section{Conclusion}
	Our measurement technique is opening a way of directly measuring very small 
tunnel splittings of the order of $10^{-8}$~K not accessible by resonance techniques. 
We have found a very clear oscillation in the tunnel splittings $\Delta$, which are direct 
evidence of the role of the topological spin phase in the spin dynamics of these molecules. 
They are also the first observation, to our knowledge, of an "Aharonov-Bohm" type of 
oscillation in a magnetic system, analogous to the oscillations as a function of external 
flux in a SQUID ring. A great deal of information is contained in these 
oscillations, both about the form of the molecular spin Hamiltonian, and also about the 
dephasing effect of the environment. We expect that these oscillations should thus become 
a very useful tool for studying  systems of quantum nanomagnets.

\section{Acknowledgment}
D. Rovai, and C. Sangregorio are acknowledged for 
help by sample preparation. 
We are indebted to P. Stamp, I. Tupitsyn, N. Prokof'ev and J. Villain for many 
fruitful and motivating discussions. 
We thank R. Ballou, A.-L. Barra, B. Barbara, A. Benoit, E. Bonet Orozco, I. Chiorescu, P. Pannetier, C. Paulsen, C. Thirion, , and V. Villar .


\begin{references}


\bibitem{Sessoli93} R. Sessoli, D. Gatteschi, A. Caneschi, and M.A. Novak, Nature (London) {\bf 365}, 141 (1993).

\bibitem{Barra96} A.-L. Barra, P. Debrunner, D. Gatteschi, Ch. E. Schulz and R. Sessoli, Europhys. Lett. {\bf 35}, 133 (1996).

\bibitem{Sangregorio97} C. Sangregorio, T. Ohm, C. Paulsen, R. Sessoli., D. Gatteschi, Phys. Rev. Lett. {\bf 78}, 4645 (1997).

\bibitem{Science} W.Wernsdorfer and R. Sessoli, Science {\bf 284}, 133 (1999).

\bibitem{Berry84} M. V. Berry, Proc. R. Soc. London A {\bf 392}, 45 (1984).

\bibitem{Haldane88} F.D.M.Haldane, Phys. Rev. Lett. {\bf 50}, 1153 (1983); Phys. Rev. Lett. {\bf 61}, 1029 (1988); see also E. Manousakis, Rev. Mod. Phys. {\bf 63}, 1 (1991), for a review (particularly section IV.E).

\bibitem{Garg93} A. Garg, Europhys. Lett. {\bf 22}, 205 (1993).

\bibitem{others} J.L. Van Hemmen and S. SŸto, Europhys. Lett. {\bf 1}, 481 (1986); Bogachek and Krive, Phys. Rev. B {\bf 46}, 14559 (1992); V.Yu. Golyshev, A.F. Popkov, Europhys. Lett, {\bf 29}, 327 (1995); I. Tupitsyn, N.V. Prokof'ev, P.C.E. Stamp, Int J Mod Phys B 11, 2901 (1997); A. Garg, Phys. Rev. Lett. {\bf 83}, 1513 (1999).

\bibitem{Loss92} D. Loss, D.P. DiVincenzo, and G. Grinstein, Phys. Rev. Lett. {\bf 69}, 3232 (1992)

\bibitem{Delft92} J. von Delft and C. L. Hendey, Phys. Rev. Lett. {\bf 69}, 3236 (1992)

\bibitem{Barnes} S.E. Barnes, J. Phys. Cond. Matter 10, L665 (1998); and cond-
mat/9907257

\bibitem{WW_PhD} W. Wernsdorfer, thesis, Joseph Fourier University, Grenoble, 
(1996); W. Wernsdorfer et al., Phys. Rev. Lett. 78, 1791 (1997).

\bibitem{remark_heat} Below 0.4 K and for $|H| >$ 0.4 T a small 
temperature dependence is observed. This is due to phonon emission 
after tunneling (see also \cite{remark4}).

\bibitem{Wieghardt} Fresh crystals were prepared according to K. Wieghardt, K. Pohl, I. 
Jibril, G. Huttner, Angew. Chem. Int. Ed. Engl. 23, 77 (1984).

\bibitem{Ohm98} T. Ohm, C.~Sangregorio, C.~Paulsen, Euro. Phys. J. B 
{\bf 6}, 195 (1998); T.~Ohm, C.~Sangregorio, C.~Paulsen, J. Low Temp. 
Phys. {\bf 113}, 1141 (1998).

\bibitem{PRL_dig} W. Wernsdorfer, T. Ohm, C. Sangregorio, R. Sessoli, D. Mailly, C. Paulsen, Phys. Rev. Lett. {\bf 82}, 3903 (1999); W. Wernsdorfer, R. Sessoli, D. Gatteschi,, Europhys. Lett. {\bf 47}, 254 (1999).

\bibitem{Zener} L. Landau, Phys. Z. Sowjetunion {\bf 2}, 46 (1932); C. Zener, Proc. R. Soc. London, Ser. A 137, 696, (1932); E.C.G. St$\ddot{u}$ckelberg, Helv. Phys. Acta {\bf 5}, 369 (1932).

\bibitem{Zener_new} S. Miyashita, J. Phys. Soc. Jpn. {\bf 64}, 3207 (1995) 
and  {\bf 65}, 2734 (1996); V.V. Dobrovitski and A.K. Zvezdin, Euro. Phys. Lett. 
38, 377 (1997); L. Gunther, Euro. Phys. Lett. 39, 1 (1997);  G.Rose, P.C.E. Stamp, J. 
Low Temp. Phys. 113, 1153 (1998); M. Thorwart, P. Reimann, P. Jung, R.F. Fox, Chem. Phys. {\bf 235} (1998) 61.

\bibitem{remark1} We supposed here that the forth and 
back sweeps give the same tunnel 
probability. This is a good approximation for $P << 1$ 
where next nearest neighbor effects can be neglected.

\bibitem{remark4} In order to avoid heating problems for measurements of $\Delta$ for $n > 1$, we started in a thermal annealed sample with $M_{\rm in} = 0.95 M_s$ instead of $M_{\rm in} = - M_s$.

\bibitem{remark5} For small sweeping rates 
(depending on $\Delta$ at high sweeping rates), the mean 
internal field changes faster than the applied field leading to strong 
deviation from the Landau Zener model.

\bibitem{remark2} The long times regime of such a plot is very sensitive to the 
equilibrium magnetization which is very difficulte to know precisely.

\bibitem{Prok_Stamp} N.V. Prokof'ev, P.C.E.~Stamp, Phys. Rev. Lett. 
{\bf 80}, 5794 (1998); N.V.~Prokof'ev, P.C.E.~Stamp, J. Low Temp. 
Phys. {\bf 104}, 143 (1996); N. V. Prokof'ev and P.C.E. Stamp, J. Phys. Condens. Matter 5, L663 (1993).

\bibitem{remark3} Recent measurements on isotope modified Fe$_8$ samples showed a 
small dependence of $\Delta$ on the hyperfine coupling which was shown in I.S. 
Tupitsyn, N.V. Prokof'ev, P.C.E. Stamp, Int J Mod Phys B 11, 2901 (1997) see eq. 3.33, and I.S. Tupitsyn, JETP Lett. {\bf 67}, 28 (1998).

\bibitem{Caciuffo} R. Caciuffo, G. Amoretti, A. Murani, R. Sessoli, A. Caneschi, and D. Gatteschi, Phys. Rev. Lett. 81, 4744 (1998).


\end{references}
\end{document}